# Modeling real spatial networks

## T. B. Progulova and B. R. Gadjiev

International University for Nature, Society and Man, Department of System Analysis and Management, 19 Universitetskaya str., Dubna, 141980, Russia

gadjiev@uni-dubna.ru

**Abstract.** We have studied transportation network, namely a road network of the Moscow region and airline network of the Russian Federation. We have constructed corresponding networks and studied degree distribution and length distribution for these networks, as well as the dependences on the average clustering coefficients and the nearest neighbours average degree as a function of the vertex degree. In conclusion we discuss degree and length distributions in the framework of the nonextensive statistics, using the maximum entropy approach and the model with additive and multiplicative noise. We present a procedure of fitting the results of the data processing to the $q$-type distribution which allows the fractal dimension definition of the networks under study.

## 1. Introduction

The system is a complex of interacting elements that form an organized entity. Any complex system can be considered as a network. In this case, the network nodes are the system elements and the edges are the interaction among the system elements. It makes it possible to represent any real physical (natural or artificial), biological and social system as a network. The essentials of the network approach are given in a number of monographs, reviews and numerous papers [1–3]. In the framework of the network approach, the system is represented as a graph. The statistical mechanics methods turned out to be convenient to study connected graphs with an unrestrictedly large number of nodes.

Studies of artificial systems show that many real networks are characterized by power-law degree distribution $p(k) \sim k^{-\gamma}$ where $k$ is the vertex degree (the number of edges incident to the given vertex) and $\gamma$ is the exponent of the distribution. It has lead to intensive studies of the evolution mechanisms of growing networks to understand topological peculiarities of networks, such as the scale-free degree distribution, the small-world property, the existence and origin of correlations in these networks [4].

Classes of networks are those embedded in the real space, i.e. networks whose nodes occupy precise positions in the two or three – dimensional Euclidian space, and whose edges are real physical connections. It is not surprising that the topology of spatial networks is strongly constrained by their geographical embedding. Transportation networks are most important among spatial networks from the social-geographical point of view [4].

Transportation networks, in one way or another, are portrayals of the history of the country's development. In any case, these networks connect places where people live and are the outcome of economic and political development of the country with an account of its geographical features. There is no doubt that historically people's settlements were chosen in accordance with a definite geographical utility of the given territory. Later, transportation networks adjusted to this structure. Transportation network in Russia mainly developed in the years of the communist rule to support the relevant social-economic policy. There was no rivalry in this system and economics was state-controlled. Transportation routes reflected the economic structure of the country which was built in such a way that the political interdependence of the republics which were its members was maximal. It should be noted that the cost of the roads was a matter of minor importance.

In case of the transportation network, there is a strong restriction to the growth of the network nodes degrees which is induced by the geographical arrangement of the edges. There is no strong geographical restriction to the edges existence in the airline network. However, the presence of connections correlates with the mean economic level of people's prosperity in relevant cities.

The topology studies of various spatial networks have showed that the latter can have significantly different degree distributions. For example, the degree distribution of the spatial Internet network (presupposing that the network nodes are routers) has the form $P(k) \sim k^{-\gamma}$ where $\gamma_{out} = \gamma_{in} \approx 2.48$ [1]. The power grid of the western US is described by a complex network whose nodes are generators transformers, and substations, and edges are high voltage transmission lines. The degree distribution of this network is expressed as exponential [5]. The analysis of the spatial distribution of the network nodes with the box counting method shows that both indicated networks are fractals [3, 6]. The algorithm that produces networks of the power grid and Internet type is given in Ref. [7]. The growing network, at definite values of the model parameters, can have both the exponential and degree form of the degree distribution [7]. The Tsallis distribution possesses this property from the point of view of the statistical mechanics approach: with the nonextensivity parameter tending to the unit, the Tsallis distribution leads to the Boltzmann-Gibbs distribution, while at large degree values it appears in the form of the power-law degree distribution. This fact makes the Tsallis statistics a convenient tool to describe the complex network topology [8–10].

Spatial networks, namely roads and airline networks are transportation networks. They solve social-economic tasks in definite geographical conditions in any specific country. A comparative analysis of them allows deeper understanding of the geographical and social-political restrictions on their operation. This procedure is necessary for a detailed study of the extension processes in these networks.

In this paper, we construct spatial road networks of the Moscow region and the airline network of the Russian Federation. We construct degree distributions of these networks. We derive the degree and length distributions in the frames of the nonextensive statistics using the maximum entropy principle and the model with additive and multiplicative noise. Applying the maximum-likelihood method, we describe the topology and define the nonextensivity parameters of the relevant networks. We assess the fractal dimensions of the road and airline networks using the dependence of the entropy index on the system fractal dimension.

**2. Data analysis**

We constructed the road network in the following way: the network nodes are road intersections and the edges are roads. There are many vertices in the constructed network with degree $k=1$, which is the result of the fact that some roads in the settlements are dead-end. Vertices with degree two are absent in the road network, as we do not account for the cities located on the roads as the network vertices. It should be noted that the account of such vertices does not make considerable changes in the road network topology. The number of vertices in the Moscow region road network is $N = 1391$, the number of edges is 1951, and the mean degree is $\langle k \rangle = 2.74$. The maximum degree of the Moscow region node equals to 6. The degree distribution of the Moscow region road network is given in figure 1.

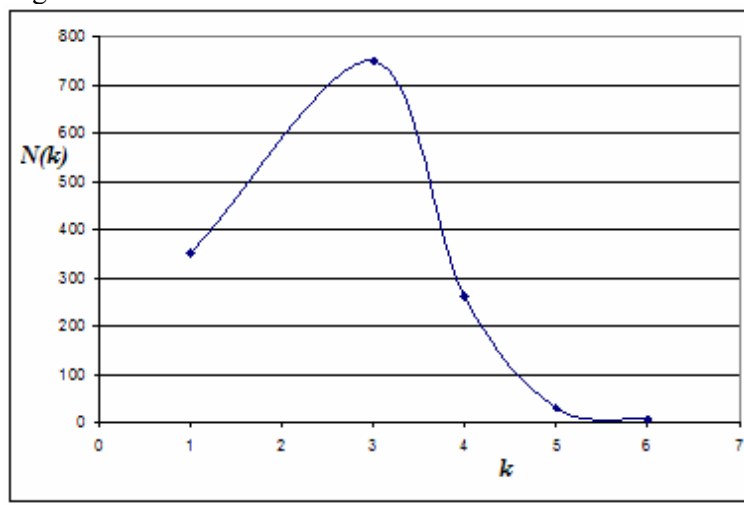

**Figure 1.** Degree distribution of the Moscow region road network.

We also architected the length distribution of the Moscow region road network. It is shown in figure 2.

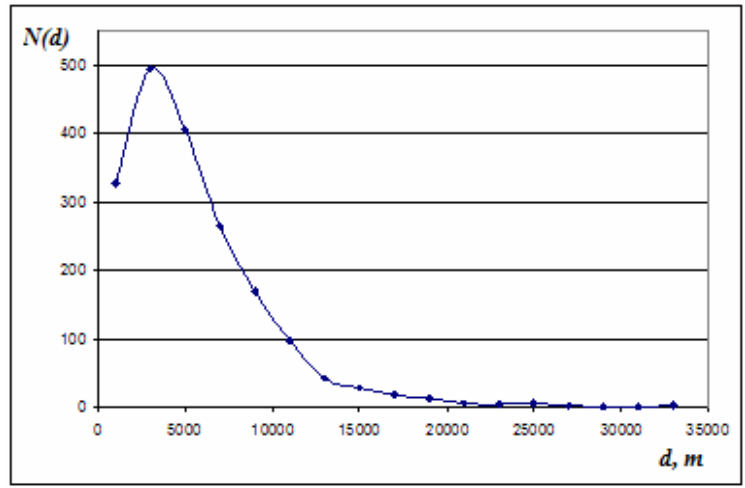
**Figure 2.** Length distribution of the Moscow region road network.

We have constructed the airline network of the Russian Federation. The network nodes are cities where airports are located, and edges stand for the availability of direct flights between relevant airports. The airline network of the Russian Federation has 190 nodes and 710 edges. Degree distribution of the RF airline network is presented in figure 3. The maximum degree of a node in the RF airline network is 239, the mean degree has the value 9.3.

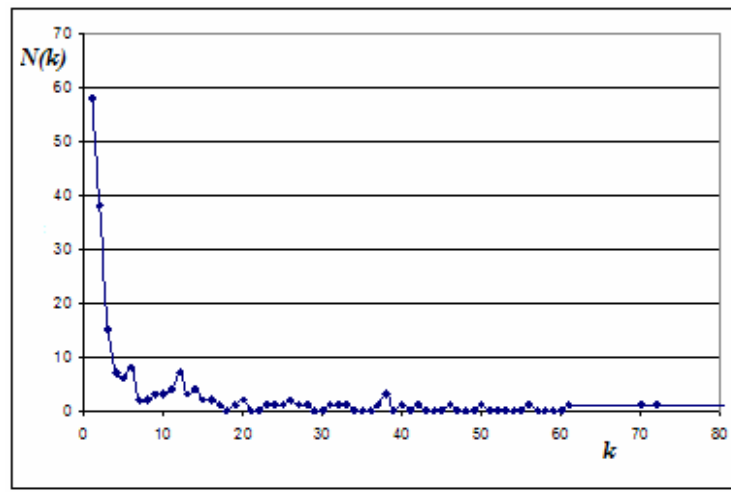
**Figure 3.** Degree distribution of the RF airline network.

We have constructed the length distribution of the airline network edges as well and give it in figure 4.

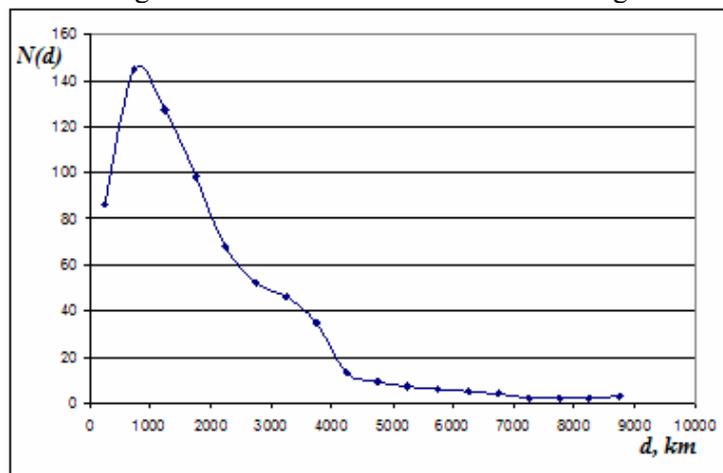
**Figure 4.** Length distribution of the RF airline network edges.

The degree distribution determines the statistical properties of uncorrelated networks. Nevertheless, a large number of real networks are correlated in the sense that the probability of a node with degree $k$ being connected to another node with degree $k'$ depends on $k$. In particular, correlated networks are classified as assortative if the nearest neighbours average degree $k_{nn}(k)$ is the $k$ increasing function, and disassortative if $k_{nn}(k)$ is the $k$ decreasing function [10]. We have constructed the dependence of the average degree of the nearest neighbours $k_{nn}$ on the $k$ nodes degrees for the road and airline networks. The results are presented in figures 5 and 6, respectively.

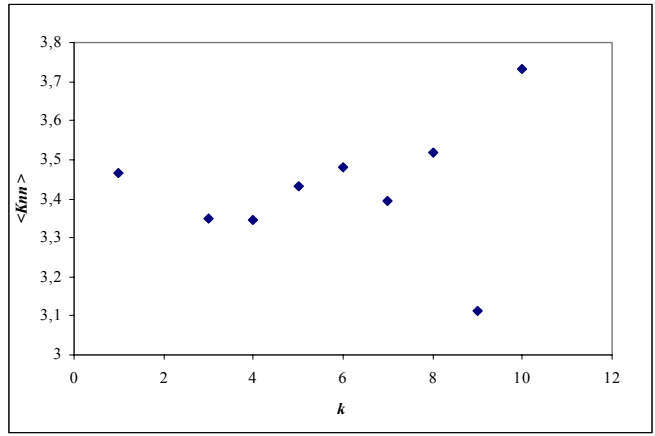

**Figure 5.** Nearest neighbours average degree $k_{nn}(k)$ as a function of the vertex degree for the Moscow region road network.

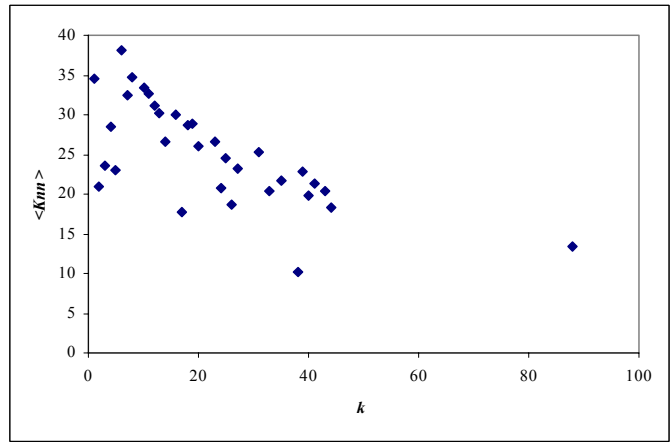

**Figure 6.** Nearest neighbours average degree $k_{nn}(k)$ as a function of the vertex degree for the RF airline network

Spatial networks generally reveal trivial dependences of the average clustering coefficient on the degree. For example, the Internet on the router level and the power grid of the western US have $c(k)$ that does not depend on $k$ [1, 4].

We have constructed the dependence of the average clustering coefficient on the degree for the road and airline networks. The results are given in figure 7 and 8, respectively. The analysis shows that the average clustering coefficient does not practically depend on the vertex degree.

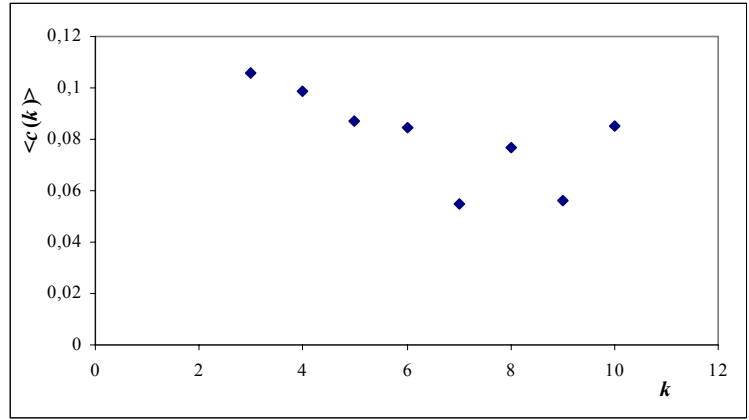

**Figure 7.** Average clustering coefficient as function of the degree *k* for the Moscow region road network.

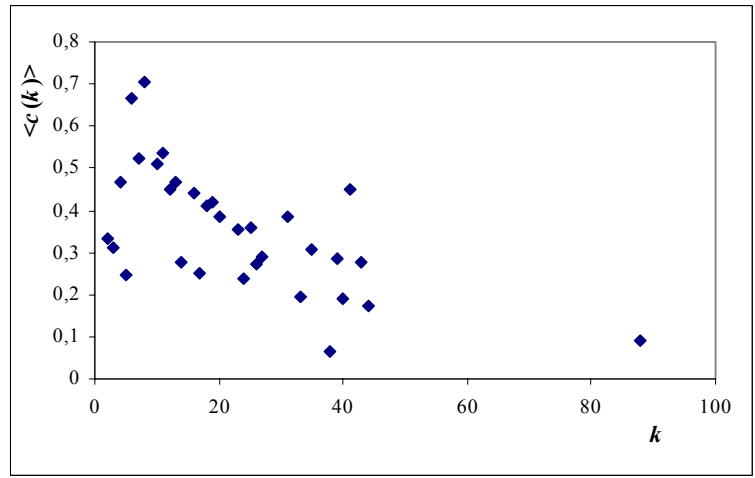

**Figure 8.** Average clustering coefficient as function of the degree *k* for the RF airline network.

The city of Moscow possesses the maximum degree and betweenness centrality for the airline network. In both networks there is no community structure and they both are not of the small-world property.

**3. Nonextensive aspects of stochastic networks**
Applying the maximum entropy principle, we can define the degree distribution function. We can introduce the entropy $S_q$ of the discussed fractal system in the form [8–10]

$$S_q = \frac{\sum_k p_k^q - 1}{1-q}, \qquad (1)$$

where $q$ stands for the entropy index of the probability distribution $p(k)$ of the $k$ states.

Natural constraints for the entropy maximization (1) are

$$\sum_k p_k = 1, \qquad (2)$$

That corresponds to the probability normalization condition,

$$\sum_k P(k)k = \mu \qquad (3)$$

$$\sum_k P(k)k^2 = \rho^2, \qquad (4)$$

where $\mu$ and $\rho^2$ are the first and the second moments of distribution $P(k)$ and

$$P(k) = \frac{p_k^q}{c_q}, \qquad (5)$$

$$c(q) = \sum_k p_k^q, \qquad (6)$$

$P(k)$ is the escort probability expressing the degree distribution.

We obtain from the variation problem for (1), taking into account restrictions (2), (3) and (4)

$$\frac{\delta}{\delta p_k}\left(\frac{\sum_{k'} p_{k'}^q - 1}{1-q} - \beta \sum_{k'} P(k')k' - \gamma \sum_{k'} P(k')k'^2 - \lambda \sum_{k'} p_{k'}\right) = 0, \qquad (7)$$

where $\alpha$, $\gamma$ and $\beta$ are the Lagrange parameters. From equation (7) we conclude that

$$p_k = \left(\frac{\lambda(1-q)}{q}\right)^{\frac{1}{q-1}} \left(1 - \frac{(1-q)}{c_q}\left(\beta(k-\mu) + \gamma(k^2 - \rho^2)\right)\right)^{\frac{1}{1-q}} \qquad (8)$$

Using the normalization requirement and introducing the notation $Z = \sum_k (1 - \xi(1-q)(k-\eta)^2)^{\frac{1}{1-q}}$, we can produce

$$p_k = \frac{1}{Z}(1 - \xi(1-q)(k-\eta)^2)^{\frac{1}{1-q}} \qquad (9)$$

Taking into account that $P(k) = \dfrac{p_k^q}{c_q} = \dfrac{p_k^q}{\sum_k p_k^q}$

We finally obtain:

$$P(k) = \frac{(1 - \xi(1-q)(k-\eta)^2)^{\frac{q}{1-q}}}{\sum_k (1 - \xi(1-q)(k-\eta)^2)^{\frac{q}{1-q}}} \qquad (10)$$

The parameters $\eta$ and $\xi$ introduced instead of the Lagrange parameters $\beta$ and $\gamma$ are defined with restrictions (3) and (4).

We can easily see that at $q \to 1$ the $P(k)$ distribution is reduced to the normal $P(k) \sim e^{-\frac{(k-\mu)^2}{2\sigma^2}}$ distribution with the mean $\mu$ value and $\sigma^2 = \rho^2 - \mu^2$ variation. We can observe from the definition of the $P(k)$ distribution that at large $k$ degrees the probability distribution has the power law form $P(k) \sim (k-\mu)^{-\frac{2q}{q-1}}$ for $q > 1$. Therefore, the $P(k)$ degree distribution varies from the exponential to power law form [10].

The maximum entropy principle with the Tsallis entropy can be applied to solve the random walk problem [12, 13]. Let $\vec{x}$ be a continuous vector variable in the $d$ — dimension of space and $p(x)$ be its probability distribution that complies with the normalization requirement. The appropriate Tsallis entropy is given in the form

$$S_q[p(x)] = -\frac{1 - \int p(x)^q dx}{1-q} \qquad (11)$$

The generalized restrictions in the $x^2$ terms are

$$\int (x - \bar{x})^2 p(x)^q dx = \sigma_q^2 d \qquad (12)$$

If we use the maximum entropy principle $S_q[p(x)]$, we can obtain the probability distribution for the value $x$, which is analogous to (10), that can be applied to the description of edges length distribution in the network.

It is convenient to discuss the mechanisms that bring about nonextensive statistics reviewing the process with both the additive and the multiplicative noises [14]. We review the stochastic equation in the form

$$\frac{du(t)}{dt} = f(u) + g(u)\xi(t) + \eta(t), \tag{13}$$

where $u(t)$ is a stochastic variable, $f, g$ are arbitrary functions and $\eta(t), \xi(t)$ are uncorrelated and Gaussian – distributed zero–mean white noises. The Fokker-Plank equation in the Stratonovich calculus appears in the form

$$\frac{\partial P(u;t)}{\partial t} = -\frac{\partial}{\partial u}(f(u)P(u;t)) + M\frac{\partial}{\partial u}\left(g(u)\frac{\partial}{\partial u}[f(u)P(u;t)]\right) + A\frac{\partial^2 P(u;t)}{\partial u^2} \tag{14}$$

where $M, A > 0$ are amplitudes of the multiplicative and additive noises, respectively. We assume that

$$f(u) = -\tau g(u)\frac{\partial g(u)}{\partial u} \tag{15}$$

where $\tau$ is the proportionality coefficient. The stationary solution of the Fokker-Plank equation is expressed in the form

$$P_{st}(u) \sim \left[1 + (q-1)\beta[g(u)]^2\right]^{\frac{1}{1-q}}, \tag{16}$$

where $\beta = \frac{\tau + M}{2A}$ и $q = \frac{\tau + 3M}{\tau + M}$. If we had used the Ito convention we would have obtained $q = \frac{\tau + 4M}{\tau + 2M}$.

In paper [15] was discussed the simulation algorithms for networks generation that provide the Tsallis distribution. It is obvious that various mechanisms may lead to the Tsallis distribution. This process, from the point of view of obtaining the degree distribution, should include evolution with preferential attachment, the formation of connections among «old» nodes and two-node merging and another mechanism of evolution. We have used the Tsallis distribution of type (10) or (16) to fit and define the system fractal dimension.

## 4. Discussion

In our case, the process of the road growth can be represented in the following way. Historically, the choice for the settlement location was determined in accordance with the economic, geographical and military attractiveness (utility). Studies of the network structure show that roads were mainly built to provide connections that linked towns and cities with the capital. The cost of the roads themselves in the road network growth was a matter of minor importance. The absence of the community structure in these networks might be explained by these circumstances. It is possibly connected to the peculiarities of the political rule history of the country, namely, one-centre administration and control. To some extent, it may also be found in the picture of the airline network. Here, the nodes with the maximal degree are at the same time those with the maximal betweenness centrality.

The analysis results, given in figure 5 and figure 6, show that the road and airline networks are uncorrelated networks. These networks are of random character described by the $q$ type distribution (14). We should indicate here that the degree distribution of the road network is close the normal one, while the degree distribution of the airline network deviates little from the power law distribution.

It is well-known that the maximum-likelihood method provides more accurate and invariable estimations [16, 17]. For this reason, we have applied the maximum-likelihood method for the empirical data fitting on the degree distribution of the road and airline networks, using (14). The results on the length fitting obtained with the maximum-likelihood method show that the vertices locations in these networks have fractal dimension $\gamma_{air} = 2.6$ и $\gamma_{moto} = 1.9$, respectively.